\documentclass{iopjournal}

\usepackage{amsmath,amsfonts}
\usepackage[
  style=authoryear,
  backend=biber,
  maxcitenames=1,
  mincitenames=1,
  maxbibnames=999,
  date=year,isbn=false,url=false,doi=false
]{biblatex}

\DeclareFieldFormat{parens}{#1}
\DeclareFieldFormat{year}{#1}

\DeclareFieldFormat
[article,inbook,incollection,inproceedings,patent,thesis,unpublished]
  {title}{#1}

\DeclareNameAlias{sortname}{family-given}
\DeclareNameAlias{author}{family-given}
\DeclareNameAlias{editor}{family-given}

\DeclareNameFormat{family-given}{%
  \usebibmacro{name:family-given}
    {\namepartfamily}
    {\namepartgiveni}
    {\namepartprefix}
    {\namepartsuffix}
  \usebibmacro{name:andothers}}
\renewcommand*{\mkbibnamegiven}[1]{#1}
\renewcommand*{\mkbibnameprefix}[1]{#1} 

\renewbibmacro*{name:family-given}[4]{%
  \usebibmacro{name:delim}{#1#2}%
  \usebibmacro{name:hook}{#1#2}%
  \mkbibnamefamily{#1}\isdot
  \ifblank{#2}{}{\bibnamedelimd\mkbibnamegiven{#2}}%
  \ifblank{#3}{}{\bibnamedelimd\mkbibnameprefix{#3}}%
  \ifblank{#4}{}{\bibnamedelimd\mkbibnamesuffix{#4}}%
}

\DefineBibliographyStrings{english}{in = {}}

\newcommand{\etal}{\textit{et~al}}
\newcommand{\bm}{\boldsymbol}

\begin{document}

\articletype{Paper} 

\title{Super-Resolution Positron Emission Tomography by Intensity Modulation: Proof of Concept}

\author{Youdong Lang$^1$\orcid{0009-0008-3432-4050}, Qingguo Xie$^{1,2,3}$\orcid{0000-0001-5353-2201} and Chien-Min Kao$^{4,*}$\orcid{0000-0002-8785-5225}}

\affil{$^1$ Biomedical Engineering Department, Institute of Artificial Intelligence, Huazhong University of Science and Technology, Wuhan, China}

\affil{$^2$ Wuhan National Laboratory for Optoelectronics, Wuhan, China}

\affil{$^3$ Department of Electronic Engineering and Information Science, University of Science and Technology of China, Hefei, China}

\affil{$^4$ Department of Radiology, The University of Chicago, Chicago, USA}

\affil{$^*$Author to whom any correspondence should be addressed.}

\email{c-kao@uchicago.edu}


\begin{abstract}
\textit{Objective}.
Spatial resolution is a critical consideration in imaging. To increase the resolution of clinical positron emission tomography (PET) systems beyond its instrumentation limit, we propose a new modulation-based approach inspired by super-resolution (SR) structured illumination microscopy (SIM), a method originally developed to overcome the diffraction-limited resolution in optical microscopy.
\textit{Approach}.
We implemented the key idea underlying SR-SIM by using a rotating intensity modulator in front of a stationary PET detector ring. Its function is to modulate down the originally unobserveable high-frequency signals of the projection data that are above the instrumentation bandwidth of a PET system to appear as aliased lower-frequency ones that are detectable. We formulated a model that relates an image whose resolution is above the instrumentation limit to several thus obtained limited-resolution measurements at various rotational positions of the modulator. We implemented an ordered-subsets expectation-maximization algorithm (OSEM) for solving the model to produce SR images.
\textit{Main results}.
We observed that theoretically SR-SIM can be applied to undersampled data produced by PET to both eliminate aliasing errors and increase resolution and confirmed this new insight numerically.
Using noise-free data produced by an analytic projector, we showed that the proposed approach can resolve 0.9\,mm sources when applied to a PET system that employs 4.2\,mm-width detectors. 
With noisy data, the SR performance remains promising. 
In particular, 1.5~mm sources were resolvable, and the visibility and quantification of small sources and fine structures were improved despite the sensitivity loss incurred by the modulator. 
Aliasings were not observed in SR images but were present in non-SR images.
These observations remain valid when using more realistic Monte-Carlo simulation data.
\textit{Significance}.
By considering specific modulator designs and using an OSEM algorithm, this paper demonstrates a practical modulation-based approach that can significantly increase the spatial resolution of a clinical PET system when higher resolution is needed.
Further systemic studies are needed for optimization of the modulator design and reconstruction algorithms. 

\end{abstract}

\section{Introduction}
\label{sec:introduction}
Positron emission tomography (PET) has demonstrated value in the diagnosis and evaluation of treatment-responses of cancer and cardiovascular and cerebrovascular diseases~\parencite{Brady2008theclinical}.
As dictated by its predominant clinical application, which is whole-body imaging for cancer diagnosis, the resolution of the present-day clinical PET systems is approximately 4-6 mm~\parencite{%
Slomka2016recentadvances,GonzalezMontoro2020Advancesindetector}.
This resolution is however inadequate for resolving small lesions and fine structures in several important organs, notably the brain.
Physical factors affecting resolution in PET include positron range, photon acolinearity and detector size.
For clinical imaging, fluorine-18 (F-18) based tracers are most widely used.
The positron range of F-18 in tissue is less than 1~mm. 
To first order, the resolution due to photon acolinearity is proportional to the scanner diameter $D$, yielding a value of about 1.8\,mm for $D=80$\,cm~\parencite{%
Shibuya_2007Acolinearity}.
Therefore, detector size is the predominant resolution-limiting factor for present PET clinical imaging and application-specific (AS) systems that employ small detectors to achieve high resolution for specific applications have been, or are being, developed~\parencite{%
    Catana2019developmentofdedicatedbrainPET,
    Gonzalez2018organdedicated,
    Ganizares2020pilotperformance}.
Nonetheless, given that clinical PET systems are readily available, it is beneficial to be able to boost their resolution on-demand for imaging certain organs or regions at higher resolution, even though higher imaging dose or longer can time may be necessary.

Improving the resolution of images derived from data acquired by clinical PET systems has been a research topic of significance~\parencite{%
Reader2007advancesinPET}.
Almost all PET scanners assume a cylindrical geometry and they are known to 
produce data that are undersampled at the center, causing aliasing errors.
Derenzo~\etal~\parencite{Derenzo1988apositrontomograph} and Suk~\etal~\parencite{Suk2008improvementofthespatial} proposed moving the detectors or the subject for improved sampling and elimination  of aliasing errors.
The so-called point-spread-function (PSF) or modeled-based (MB) image reconstruction methods model the detector response and employ it in reconstruction for removing its resolution-limiting effects up to the Nyquist frequency of sampling or the bandwidth of the detector response~\parencite{Rahmim2013resolutionmodeling}.
When applying MB methods to oversampled data produced by using detector/subject motions or by using detectors capable of measuring depth-of-interaction (DOI)~\parencite{%
jeong_sinogram-based_2011, Zang-Hee2019wobbleandzoom, kao_2002},  resolution and resolution uniformity can be improved further.
Metzler~\etal~\parencite{%
Metzler2013resolutionenhance}
proposed to effectively reduce the detector size  by collimation,
hence increasing the detector resolution, in conjunction with using MB reconstruction and employing specific collimator movements to eliminate undersampling.
Although collimation reduced the coincidence detection efficiency (CDE) by approximately 75\%, they reported better visualization and quantification for small structures.
For boosting resolution locally, the idea of using small insert detectors for producing higher-resolution measurements for a region-of-interest has been proposed~\parencite{%
    Tai2008virtualpinhole,
    jiang2019asecondgeneration,
    Zhou2009theoreticalanalysis}.

Recently, various super-resolution (SR) techniques for surpassing the inherent resolution limit in microscopy due to diffraction of light have been demonstrated~\parencite{%
Schermelleh2019superresolutionmicroscopydemystified}.
In structured illumination microscopy (SIM)~\parencite{%
    Allen2014structuredilluminatinmicroscopy,
Gustafsson2005nonlinearstructuredillumination},
the key idea is to make originally undetectable signals that are above the bandwidth of an imaging device detectable by modulating them down to appear as lower-frequency ones that lie within the bandwidth, albeit as aliased signals. The modulation pattern is translated to produce a number of limited-resolution measurements with different modulations. Then, an SR image is produced from these measurements by computationally de-aliasing the signals. This work is based on the observation that the modulation step in SR-SIM can be achieved in PET by introducing a rotating intensity modulator in front of a stationary PET detector ring, and on the postulation that the continuous, shift-invariant SR-SIM theory can be extended to the discrete, shift-variant measurement system of PET.
It is worth noting that this SIM-based approach and the collimation-based method by Metzler~\etal\,
are based on two distinct theoretical perspectives.
Firstly, as will be discussed in section~\ref{subsec:fundamental}, SR-SIM in theory can expand the resolution bandwidth of an imaging device infinitely~\parencite{%
    Gustafsson2005nonlinearstructuredillumination}. 
This is not the case with the collimation-based method.
Secondly, compared to collimators, an intensity modulator incurs less reduction of CDE.
Also, it is easier to implement modulator rotation than the nontrivial collimator movements proposed Metzler~\etal.

The objective of this paper is
to demonstrate this SR-SIM inspired approach for achieving super-resolution PET (SR-PET)
by using simulation data.
As a proof-of-concept, we will model a two-dimensional (2-d) system 
and neglect attenuation by subject, scatter, randoms, variation in detector efficiency, positron range, photon acolinearity and DOI blurring.
As in all signal-recovery problems, performance of the approach will be affected by the characteristics of the measurement model, the algorithm used to solve the model, and data statistics.
Based on using an ordered-subsets expectation-maximization (OSEM) algorithm, we will examine the impacts of data noise and certain design parameters of the modulator.
Comprehensive examination of the effects of all resolution-affecting factors and optimization of the modulator design and reconstruction algorithm will be considered in subsequent studies.

\section{Background and Theory}
\label{sec:background}

Below, we will use $x$ to represent the spatial coordinate and $\nu$ the associated spatial frequency. We will use lowercase letters such as $a(x)$ to denote spatial functions and uppercase letters such as $A(\nu)$ to denote the associated Fourier transforms (FT).
A function $a(x)$ is said to be $\nu_B$-bandlimited if $A(\nu)=0$ for $|\nu|\geq \nu_B>0$ where $\nu_B$ is the bandwidth (BW).
We consider real signals and systems; therefore, it suffices to consider only positive frequencies. 
The complex conjugate of a number $a$ is denoted by $a^*$. A function and its FT will be used interchangably.

\subsection{Fundamentals of resolution recovery}
\label{subsec:fundamental}

We consider a one-dimensional (1-d) linear-shift invariant (LSI) system
whose response is characterized by the PSF $h(x)$ so that the output of the system for an input signal \(f(x)\) is given by \(g(x)=h(x)\star f(x)\), where \(\star\) denotes the convolution operator, or \(G(\nu)=H(\nu)F(\nu)\) in the Fourier space.
We assume that the LSI system has a finite BW $\nu_B$; that is,
$H(\nu)$ is $\nu_B$-bandlimited.
Evidently, \(G(\nu)=0\) for \(|\nu|\geq\nu_B\) and the measurement carries no information about the input for frequencies above $\nu_B$.
In practice, \(H(\nu)\) typically rolls off from the maximum value \(H(0)\) at $\nu=0$ to identically zero at $\nu_B$.
Hence, higher-frequency components of \(f(x)\) within the BW are also diminished, further reducing the resolution.
In theory, \(f(x)\) can be recovered up to $\nu_B$ by computing the pseudoinverse $f^\dagger(x)$ given by $F^\dagger(\nu)=H^\dagger(\nu)G(\nu)$ where $H^\dagger(\nu)=1/H(\nu)$ for  $|\nu|<\nu_B$ and $H^\dagger(\nu)=0$ for  $|\nu|\geq\nu_B$.
If the measurement is sampled, the Nyquist frequency $\nu_N=(2\delta x)^{-1}$, where $\delta x$ is the sampling distance, must exceed $\nu_B$ to avoid aliasing.
Otherwise, resolution recovery is limited to frequencies where aliasing is negligible; these frequencies can be much lower than $\nu_N$, which in turn is less than $\nu_B$ by hypothesis.

In view of this simplified theory, the above-discussed MB image reconstruction methods for PET are equivalent to computing the pseudoinverse solution and, if needed,
employing oversampled data to eliminate aliasing and allow full restoration of signal up to $\nu_N$.
The work of Metzler~\etal\ expands the BW of a PET system by effectively
decreasing the detector size while increasing the sampling to the level
that is commensurate with the expanded BW.

\subsection{Basic theory of SR-SIM}
\label{subsec:SIM-theory}

SR-SIM is based on a different theory than discussed above for achieving BW expansion.
For readers' convenience,
a concise overview of the 1-d LSI theory of SR-SIM is given below.
Readers can consult \parencite{%
Schermelleh2019superresolutionmicroscopydemystified,
Allen2014structuredilluminatinmicroscopy,
Gustafsson2005nonlinearstructuredillumination}
for in-depth discussion.

In SR-SIM, the signal is modulated by a translated, periodic signal $m(x)$ of frequency $\nu_0$ before measurement, yielding \begin{equation}
    g_\ell(x)=h(x)\star\{m(x-x_\ell)f(x)\},
    \label{eq:SIM-measurement}
\end{equation}
where $x_\ell$,  $\ell=1,\cdots,L$, is some translation.
According to the Fourier series theory, we can write
\begin{equation}
    m(x)=\sum_{k=-K}^{K} m_k\, e^{j2\pi k\nu_0x}
    \text{\ \ with\ \ } m_{-k}=m^*_{k},
    \label{eq:modulation}
\end{equation}
where we have assumed $m_k=0$ for $k>K$. Using equation~\eqref{eq:modulation} in equation~\eqref{eq:SIM-measurement} and taking FT, we get
\begin{equation}
    G_\ell(\nu) =  \sum_{k=-K}^K c_{\ell k}\,F_k(\nu)
    \label{eq:SIM-freq-1}
\end{equation}
where $c_{\ell k}=m_ke^{-j2\pi k\nu_0x_\ell}$ and
\begin{equation}
    F_k(\nu)=H(\nu)F(\nu-k\nu_0).
    \label{eq:SIM-freq-2}
\end{equation}
Evidently, $F_k(\nu)$ and $G_\ell(\nu)$ are $\nu_B$-bandlimited due to $H(\nu)$.
If there are $L=2K+1$ measurements and the resulting $(2K+1)\times (2K+1)$ matrix $\mathcal{C}=[c_{\ell k}]$ is invertible, using equation~\eqref{eq:SIM-freq-1} one can solve $\{F_k(\nu)\}_{k=-K}^{K}$ from $\{G_\ell(\nu)\}_{\ell=-K}^{K}$ for each frequency $|\nu|<\nu_B$.
Then, by computing $F(\nu-k\nu_0) = H^\dagger(\nu) F_k(\nu)$
for $|\nu|<\nu_B$, one obtains $F(\nu)$ for $\nu\in\Omega_k=(k\nu_0-\nu_B,k\nu_0+\nu_B).$ If $k\nu_0+\nu_B > (k+1)\nu_0-\nu_B$, or
\begin{equation}
    \nu_0 < 2\nu_B,
\end{equation}
then adjacent $\Omega_k$'s overlap and $F(\nu)$ can be recovered up to the expanded BW given by
\begin{equation}
    \nu_B'=K\nu_0+\nu_B. \label{eq:expanded-BW-classical_SRSIM}
\end{equation}

Therefore, by using an appropriate modulation $m(x)$ and a number of translations to yield an invertible $\mathcal{C}$, the BW of the system can be effectively expanded at the cost of needing multiple measurements. Generally, more measurements are needed for achieving greater BW expansion. In theory, as there is no limit on $K$, the expansion can be infinite. In reality, the achievable expansion depends on the numerical properties in inversion of $\mathcal{C}$, and hence on the inversion algorithm,
data statistics, and modeling accuracy.

\subsection{SR-SIM applied to sampled measurements}
\label{subsect:sampled-measurements}

In practice, samples $g_\ell(n\delta x)$ of $g_\ell(x)$ are obtained
with some sampling interval $\delta x>0$.
Using these samples, one can construct the sampling function
\begin{equation}
g_\ell^s(x)=\sum_{n=-\infty}^\infty g_\ell(n\delta x)\delta(x-n\delta x).
\end{equation}
According to the Shannon sampling theory~\parencite{%
    Barrett2004Foundationsofimagescience}
and using equation~\eqref{eq:SIM-freq-1},
the FT of $g_\ell^s(x)$ is given by
\begin{equation}
    G_\ell^s(\nu) = \sum_{n=-\infty}^\infty
    G_\ell(\nu-n\nu_s) = \sum_{k=-K}^{K} c_{\ell k} F^s_k(\nu),
    \label{eq:SIM-freq-aliasing-1}
\end{equation}
where $\nu_s = 1/\delta x$ is the sampling frequency and
\begin{equation}
    F^s_k(\nu)=\sum_{n=-\infty}^\infty F_k(\nu-n\nu_s)
\end{equation}
is the FT of the sampling function constructed from samples of $f(x)$.
Now, inverting equation~\eqref{eq:SIM-freq-aliasing-1}
no longer yields $F_k(\nu)$ but $F^s_k(\nu)$.
Under the condition
\begin{equation}
    \nu_s\geq 2\,\nu_B, \label{eq:sampling_condition}
\end{equation}
one can extract $F_k(\nu)$ from $F^s_k(\nu)$ by lowpass filtering and achieve the expanded BW $\nu_B'$.
In microscopy, the sampling condition given in equation~\eqref{eq:sampling_condition} is satisfied by using CCD or CMOS photosensors whose pixel size is sufficiently small with respect
to the PSF of the optical system.

But a cylindrical PET system consisting of discrete detectors
is known to yield undersampled data.
Let $d$ be the detector width and consider a projection profile near the center of the system.
The coincidence response function is approximately a triangle with a base equal to $d$~\parencite{%
    GonzalezMontoro2020Advancesindetector}.
The FT of this response function is proportional to
$(\sin(\pi\nu w)/(\pi\nu w))^2$ with $w=d/2$.
This function never becomes identically zero at above a certain frequency.
However, its absolute amplitude decays rapidly and a common practice
is to use its first zero as its BW, yielding $\nu_B=2/d$.
But $\delta x=d$. Hence, $\nu_s=1/d=\nu_B/2$, violating equation~\eqref{eq:sampling_condition}.

However, we observe that SR-SIM is applicable for undersampled data
to both eliminate aliasing errors and expand the BW.
By defining $Q_k(\nu)=F_k(\nu+k\nu_0)$ and $R_k(\nu)=H(\nu+k\nu_0)$, we can rewrite equation~\eqref{eq:SIM-freq-2} as the outputs of multiple LSI systems
with a common input $f(x)$:
\begin{equation}
    Q_k(\nu)=R_k(\nu)F(\nu). \label{eq:multiple-outputs-LSI}
\end{equation}
According to the Papoulis' generalized sampling theorem (GST)~\parencite{Papoulis1977GeneralizedSampling}, which is reviewed in the Appendix, if $f(x)$ has a BW $\sigma$, under mild conditions it can be recovered from samples of $(2K+1)$ LSI outputs $q_k(x)$ obtained using $\nu_s= 2\sigma/(2K+1)$.
Note that in general $q_k(x)$ also has a BW $\sigma$;
hence, this $\nu_s$ represents undersampling by a factor of $(2K+1)/2$.
Conversely, the BW of the recovered $f(x)$ is
\begin{equation}
    \sigma=(2K+1)\times (\nu_s/2). \label{eq:GST_BW}
\end{equation}
Now, since $F^s_k(\nu)$'s can be obtained from the sampled measurements by solving equation~\eqref{eq:SIM-freq-aliasing-1} and 
$Q^s_k(\nu)=\sum_{n=-\infty}^\infty Q_k(\nu-n\nu_s)=\sum_{n=-\infty}^\infty F_k(\nu+k\nu_0-n\nu_s)=F_k^s(\nu+k\nu_0)$, samples of $q(x)$ are also known.
The above observations therefore suggest that SR-SIM can be applied to undersampled data to recover signal up to the frequency $\sigma$ given in equation~\eqref{eq:GST_BW}.
Unlike $\nu_B'$ given in equation~\eqref{eq:expanded-BW-classical_SRSIM} for classical SR-SIM, here the expanded BW $\sigma$ depends only on $\nu_s$, not $\nu_0$.
However, we postulate that the two conditions of the GST given in the Appendix may result in certain requirements on $\nu_0$ in relation to $\nu_s$. 

\section{Material and Method}
    
\subsection{Mathematical formulation for SR-PET}
\label{subsec:SR-PET}
Extending the above SR-SIM theory to 2-d PET is based on making the following associations:
\begin{itemize}
    \item $f(x)$ is a profile of the parallel-beam projection of the unknown image in some projection angle.
    \item $m(x)$ is an intensity modulation of the projection profile achieved by, for example, introducing a periodic pattern of attenuation before the radiation beams are detected.
    \item PET data obtained in a projection angle is the sampled, blurred measurement of the modulated projection profile. 
\end{itemize}
Then, conceptually by applying the SR-SIM method an SR profile can be computed from several measured profiles obtained by using a set of translated $m(x)$.
Once SR profiles in all projection angles are obtained, an SR sinogram is obtained.
Reconstruction of the SR sinogram then yields an SR image.

In reality, the above is complicated by the circular geometry of a PET system, leading to nonstationary resolution responses and nonuniform sampling of the projection profiles.
It is also natural to consider rotation of a "ring modulator" that
consists of a periodic attenuation pattern on a ring
as such a modulator can be introduced unobstructively
in front of a stationary PET detector ring.
However, with respect to the projection profile, the resulting modulation $m(x)$ is not strictly periodic and an regularly-spaced rotation of the modulator does not lead to a regularly-spaced translation of $m(x)$.
Moreover, the sampling condition for image pixels away from the center of the scanner is more complicated as the projection profiles in different angles give different sampling densities.
Despite these departures from the model assumed in the above section,
we hypothesize the approach remains applicable.
Also, although we conceptualize achieving an SR sinogram first,
we seek to solve SR images directly from PET data.

Hence, let $\bm{f}$ represent an image, with its element $f_j$ being the value at pixel $j$, and $\bm{y}_\ell$ the acquired PET data with its elements $y_{\ell,i}$ being the event count at LOR $i$ when the modulator is at the $\ell$th rotation position. Ignoring scatter, randoms and attenuation by subject, we can write
\begin{equation}
\bm{y}_\ell = \textrm{Poisson}(\mathcal{H}_\ell \bm{f}),\ \ \ell=1,\cdots,2K+1,
\label{eq:PET-SIM-patternl}
\end{equation}
where $\mathcal{H}_\ell$ is the system response that, 
in addition to the detector response,
also includes attenuation by the modulator
and $\textrm{Poisson}(\bm{a})$ is a vector
whose elements are Poisson variates whose means
equal to the corresponding elements of $\bm{a}$.
By juxtaposing $\bm{y}_\ell$'s to yield a single data vector $\bm{Y}$,
we obtain
\begin{equation}
    \bm{Y} = \textrm{Poisson}(\mathcal{H}'\bm{f}) \ \ \text{with}\ \
    \mathcal{H}'= 
    \begin{bmatrix}
        \mathcal{H}_1 \\
        \mathcal{H}_2  \\
        \vdots
    \end{bmatrix}.
    \label{eq:PET-SIM-patternl-1}
\end{equation}
This equation, identical to the standard PET imaging equation in form,
can be solved by using any existing algorithm for PET image reconstruction,
including the popular OSEM algorithm~\parencite{Reader2007advancesinPET}
that accelerates the expectation-maximization (EM) algorithm for finding
the maximum-likelihood (ML) solution of equation~\eqref{eq:PET-SIM-patternl-1}.
Evidently, the image pixel size shall be chosen according to
the anticipated resolution of the solution image.

\subsection{PET system and modulators}
\label{subsec:system_and_modulator}

We considered a 77\,cm-diameter detector ring consisting of $576$ detectors whose width $d$ was 4.2~mm. These parameters resemble those of a present-day clinical PET system.

\begin{figure}
\centering
\includegraphics[width=0.5\linewidth]{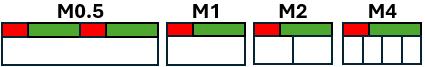}
\caption{%
Schematics of four modulators having different periods
on top of the detectors (white rectangles).
The transmission values in the red and green segments are
0.76 (emulating 5~mm thick tungsten)
and 1 (emulating air), respectively.
For purpose of visualization, these drawings are not properly scaled;
otherwise, white rectangles representing detectors shall have the same width.}
\label{fig:modulator}
\end{figure}

As shown in Figure~\ref{fig:modulator},
the modulators, placed immediately before the detectors,
were defined by repeating a basic period of which
one third having a transmission value of 0.76 and the rest having a value of 1.
We note that the value 0.76 is the probability for a 511\,keV photon
to pass 5\,mm thick tungsten without interaction.
Therefore, these represent bi-level modulators that are consisted of
alternating intervals of one-unit length of 5\,mm-thickness tungsten 
and two-unit length of clearance.
We considered four modulators, designated as M$\alpha$, whose periods were $\alpha d$ with $\alpha=0.5,1,2,4$.
For all modulators, three rotational positions (corresponding to $K=1$)
whose angular spacing is one-third of the angular width of
one basic period were used.
Hence, the angular positions of the modulator were
$\phi_\ell=\ell\,\delta\phi/3$ for $\ell=0,1,2$ where $\delta\phi=2\alpha\pi/N_d$.

When using one of the above modulators, the resulting CDE relative to the CDE in the situation when no modulator was used is $\beta(5\,\text{mm})=0.57$.
Here, the argument indicates the tungsten thickness used.
Intuitively, a bi-level modulator with a smaller $\beta$ creates a larger depth of modulation to allow for better recovery of the aliased signals.
On the other hand, it also results in larger CDE reduction and more degraded data statistics.
To examine the potential impacts of the modulation depth,
for M2 we also examined several tungsten thicknesses,
including 1\,mm, 2.5\,mm, and 10\,mm.
Their relative CDEs are
$\beta(1\,\text{mm})=0.85$,
$\beta(2.5\,\text{mm})=0.71$,
and $\beta(10\,\text{mm})=0.48$.

\subsection{Activity phantoms}
\begin{figure}
    \centering
    \includegraphics[width=0.5\linewidth]{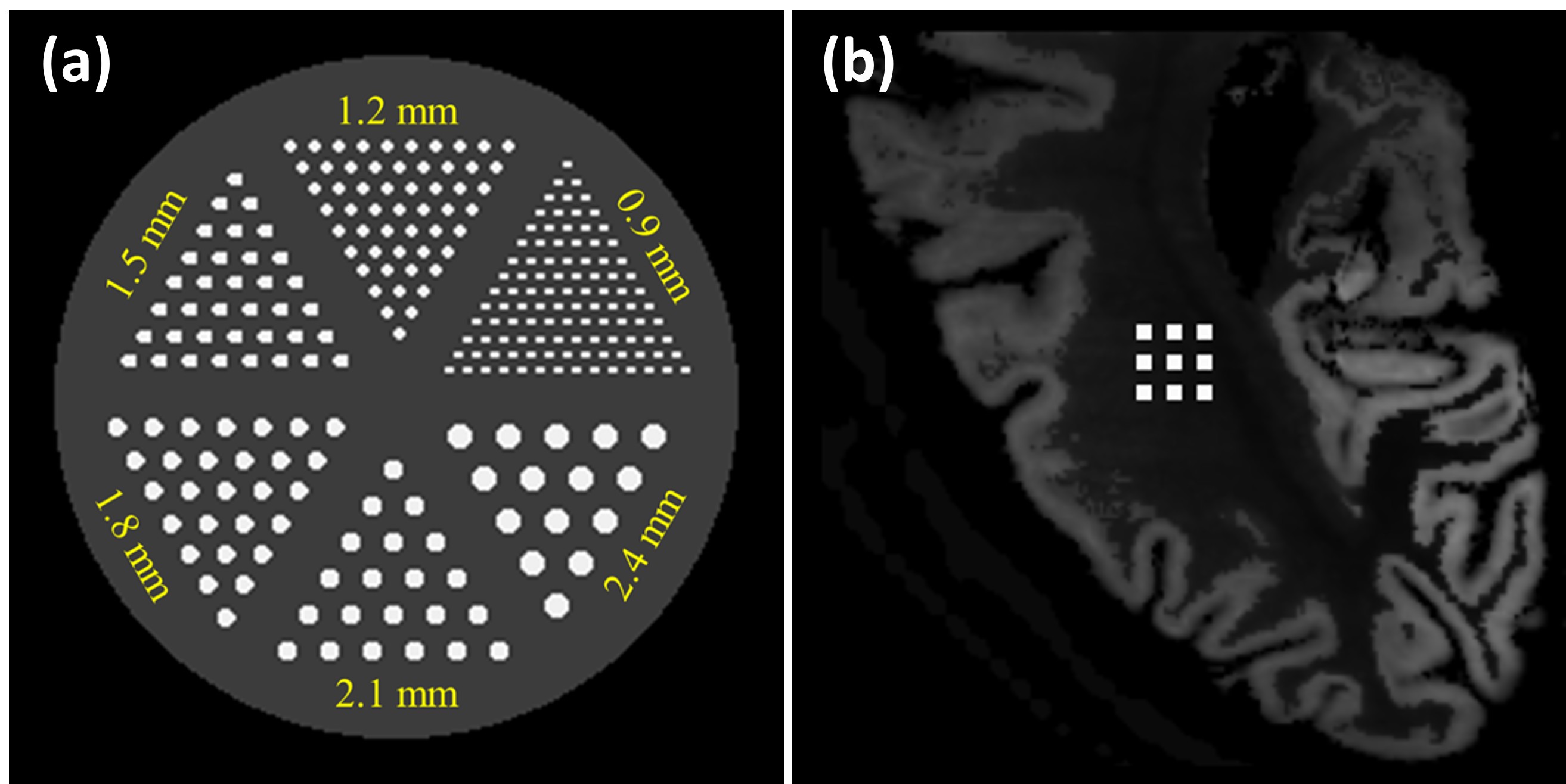}
    \caption{Activity phantoms employed. See text for details.}
    \label{fig:phantom}
\end{figure}

Figure~\ref{fig:phantom}(a) shows the activity phantom for evaluating resolution properties.
It contained six sectors of equal-diameter sources plus a 69.0~mm-diameter background disc.
The diameters of the sources in the six sectors were 0.9, 1.2, 1.5, 1.8, 2.1, and 2.4\,mm.
The center-to-center spacing between two adjacent sources in the same sector was twice their diameter.
The activity ratio of the sources to the background disc was 4:1. 
Below, the 6 sectors of this \textit{resolution phantom} are labeled numerically from 1 to 6 in increasing order of the source diameter.
For more realistic distributions, we also constructed an activity phantom by magnifying a portion of a 2-d slice of a high-resolution brain phantom
created by Belzunce \etal~\parencite{Belzunce_2020Phantom} using bilinear interpolation.
Nine 1.5\,mm square lesions were added to a region near the center.
The lesion, gray matter, and white matter activity ratio is approximately 12:3:1.
Figure~\ref{fig:phantom}(b) shows the resulting \textit{brain phantom}.
Both phantoms are a 256$\times$256 array of 0.3\,mm square pixels.

\subsection{Analytic simulation}
\label{subsec:data generation}

For accuracy, we divided a crystal into $m$ subcrystals of equal width
and defined virtual LORs (vLOR) by connecting the centers of any two subcrystals.
In front of the crystals, we assumed the presence of a zero-thickness ring modulator
that was described mathematically by a bi-level transmission profile
for 511 keV as described in section~\ref{subsec:system_and_modulator}.
Given a numerical activity phantom, for a given vLOR
we computed the raysum of the phantom $s$ by using the Siddon's algorithm~\parencite{Siddon1986fastcalculation} and
identified the transmission coefficients $t_1$ and $t_2$
of the ring modulator at the two locations where the vLOR intersects.
The modulated projection for a given crystal pair was then obtained
by summing the products $s\cdot t_1\cdot t_2$ of
the $m^{2}$ vLORs belonging to that crystal pair.
Using a larger $m$ can improve modeling accuracy but require more computation. Unless mentioned otherwise, $m=24$ was used in this work.

When a modulator was used, a noise-free dataset was generated for an activity phantom using the above-described analytical projector at each of the three rotational positions.
As described in section\ref{subsec:SR-PET}, the resulting datasets were juxtaposed to produce a single data vector $\bm{Y}$.
When no modulator was used, $\bm{Y}$ contained a single noise-free dataset.
Noisy data vectors were obtained by adding Poisson noise to the noise-free data vectors.
Noise level was specified in terms of the expected total number of events $N$ in the situation when no modulator was used.
In case when a modulator was used, the actual expected total event count was $\beta N$, where $\beta$ was the relative CDE of the modulator.
Therefore, the simulated noisy data included CDE reduction by the modulator.

\subsection{Monte-Carlo simulation}
\label{method:GATE simulation}


We also used GATE~\parencite{Sarrut_2022GATE}, which is a popular open-source Monte-Carlo (MC) simulation package for PET, for more accurate modeling of measurement.
A single-ring PET system having the 2-d geometry given in section~\ref{subsec:system_and_modulator} and an axial thickness 4.2\,mm (equal to $d$) was simulated.
The detector material was lutetium-yttrium oxyorthosilicate (LYSO).
The modulators were modeled as consisting of alternating intervals of tungsten and clearance as described in section~\ref{subsec:system_and_modulator}.
Specifically, the M2 modulator was modeled as $N_d/2=288$ tungsten blocks equally spaced on a ring. The dimension of each block was 2.8\,mm (transaxially) $\times$ 4.2\,mm (axially) $\times$ 5\,mm (radially). There is a 1\,mm gap of air between modulator and detector.
When a modulator was used, the simulated data were the collection of all events generated by three separate simulation runs with the same scan duration, each for one rotational position of the modulator.

Back-to-back 511\,keV gamma photons were produced.
To speed up simulation, photon emission angles were confined to a transaxial plane.
As already discussed, in this work we considered only  the resolution-limiting effects of the detector size while ignoring other factors including photon acollinearity, positron range, and randoms.
To minimize the number of scattered events, the energy resolution was set to 1\% and the energy window to 480-650~keV.
A 249\,ps FWHM coincidence time resolution and a 4\,ns coincidence time window were used.

\subsection{Image reconstruction}
\label{method:image-reconstruction}

For reconstruction of the analytically simulated data,
the procedure described in section~\ref{subsec:data generation} was used
to compute the forward projection of an image estimate on-the-fly.
It is easy to modify the procedure for on-the-fly computation of backward projection $\sum_i \mathcal{H}'_{ij} r_i$ as follows:
For each of the $m^2$ vLORs belonging to crystal pair $i$, the value $r_i\cdot(t_1\cdot t_2\cdot \ell)$ is add to all pixels on a vLOR, where $\ell$, $t_1$ and $t_2$ are defined above.
Like the activity phantoms, the solution images were also a 256$\times$256 array of 0.3\,mm square pixels.
For noise-free data, we employed as many OSEM iterations as practically feasible to ensure convergence.
For noisy data, the algorithm was terminated when the image was subjectively determined to have reached a good balance between resolution and noise.
No post-reconstruction image smoothing was applied.
Unless otherwise stated, 16 random subsets were used for OSEM.

For reconstruction of the MC data, the system response matrix was pre-computed using GATE, by placing 400\,kBq point sources at the centers of all image pixels.
For improved statistics and reduced system-matrix size, entries of the system matrix having the same value due to symmetry of the scanner and modulator were averaged for improved statistics. 
After averaging, small entries whose values are less than 2\% of the maximum value of all entries were truncated to zero due to their poor statistics. 
No other smoothing was applied.
Long scan durations were simulated to ensure that the number of events collected from each image pixel is no fewer than 140 thousands for M0 and no fewer than 80 thousands at each rotational position of the modulator for M2.

Below, the resulting simulated data and images will be designated as M0.5, M1, M2 or M4 according to the modulator employed. On the other hand, those obtained without using a modulator are designated as M0.
For comparison, in the case of analytic data we also reconstructed the M0 data by using analytic forward/backward projectors calculated by using $m=1$ and designated the resulting images as S.
This reflects the common practice of associating PET data with the raysums of the unknown image along LORs connecting the front centers of a pair of detectors. 
Therefore, S and M0 correspond to the present practice of performing image reconstruction without and with PSF correction, respectively. 
According to the conventional rule, the image resolution at the center of the scanner with no PSF correction is equal to $d/2$, or 2.1\,mm for the simulated system.
The Shannon sampling theory then suggests that $\sim$1.0\,mm pixels shall be used for the solution image for smaller pixels are susceptible to generation of noise and aliasing errors at frequencies where signals are negligible. 
Hence, for S images using analytic projectors, in addition to 0.3\,mm pixels we also considered 1.0\,mm pixels due to the above consideration.

 \subsection{CRC-versus-STD trade-off}
 \label{method:CRC-STD}
 For quantitative evaluation, we consider the trade-off properties between the contrast-recovery-coefficient (CRC) and the standard deviation (STD) as the number of OSEM iteration varies. Let $\mu(\bm{f},\mathcal{A})=\sum_{i\in\mathcal{A}} f_i/N_\mathcal{A}$ be the average of an image $\bm{f}$ over pixels in a set $\mathcal{A}$, where $N_\mathcal{A}$ is the number of pixels in $\mathcal{A}$, and $\sigma(\bm{f},\mathcal{A})=[\sum_{i\in\mathcal{A}} (f_i-\mu(\bm{f},\mathcal{A}))^2/(N_\mathcal{A}-1)]^{1/2}$ be the standard deviation. For sector $k$ of the resolution phantom, we define $\mathcal{S}_k$ and $\mathcal{B}_k$ to be the sets that include pixels that are inside and outside of the sources in sector $k$, respectively. Given an image $\hat{\bm f}$ obtained for the phantom ${\bm f}$, its contrast in sector $k$ is $C_k(\hat{\bm f})=\mu(\hat{\bm f},\mathcal{S}_k)/\mu(\hat{\bm f},\mathcal{B}_k)-1$. Its CRC and STD in sector $k$ are then calculated by
\begin{align}
    \text{CRC}_k&=C_k(\hat{\bm f})/C_k(\bm{f}),\label{eq:CRC}\\
    \text{STD}_k&=\sigma(\hat{\bm f},\mathcal{B}_k). \label{eq:STD}
\end{align}
CRC and STD are similarly obtained for the lesions introduced in the brain phantom. In this case, $\mathcal{S}$ for computing CRC includes pixels of the lesions and $\mathcal{B}$ for computing STD includes pixels surrounding the lesions inside a 10.5\,mm square area centered at the lesions.

\begin{figure}[ht]
    \centering
    \includegraphics[width=0.67\linewidth]{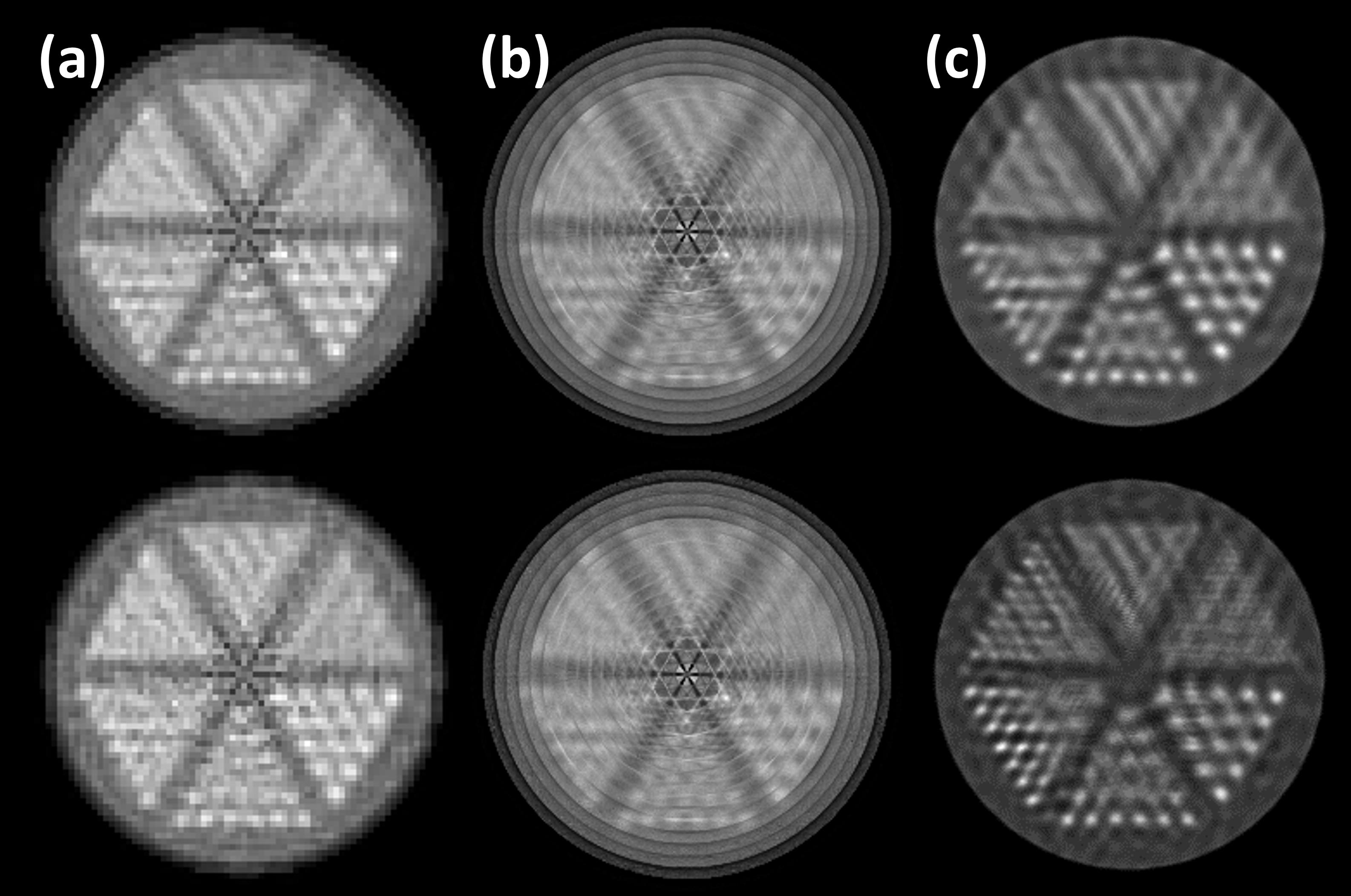}
    \caption{S image using 1.0\,mm pixels (a) and 0.3\,mm pixels (b), and M0 images using 0.3\,mm pixels (c) obtained from analytically simulated noise-free data by using 50 (top) and 500 (bottom) OSEM iterations. The phantom was placed at the center.}
    \label{fig:noise-free-center-M0}
\end{figure} 

\begin{figure}[ht]
    \centering
    \includegraphics[width=0.9\linewidth]{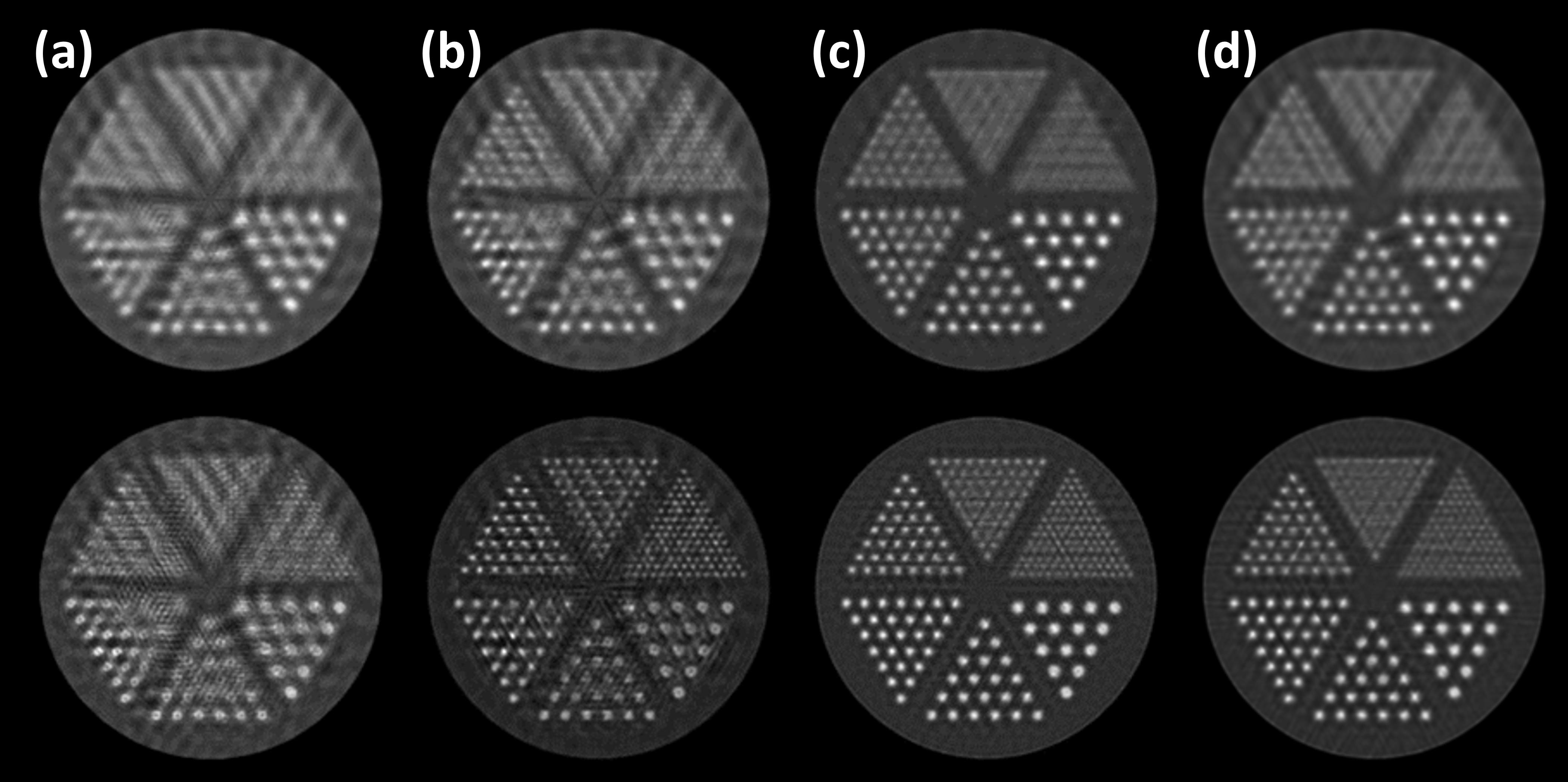}
    \caption{M0.5 (a), M1 (b), M2 (c), and M4 (d) images obtained from analytically simulated noise-free data by using 50 (top) and 500 (bottom) OSEM iterations. The phantom was placed at the center.}
    \label{fig:noise-free-center-M}
\end{figure}

\section{Result}

Due to the increased computational burden with the SR-PET approach, the activity phantoms considered were small, only 7.68\,cm$\times$7.68\,cm.
In this work, the phantoms were placed at the center of the scanner.
Thus, the results below demonstrate the SR performance properties in the central region of the scanner where sampling is nonuniform but undersampling occurs.
Evidently, DOI blurring was not modeled in analytic simulation.
MC simulation included DOI blurring; however, the effect was negligible near the center of the scanner.

\subsection{Images of the resolution phantom -- analytically simulated data}
Figure~\ref{fig:noise-free-center-M0} shows the S and M0 images obtained from analytically simulated noise-free data for the resolution phantom placed at the center of the scanner by using 50 and 500 OSEM iterations.
In the S images using 1.0\,mm pixels, 2.1\,mm and 2.4\,mm sources can be distinguished with confidence and 1.8\,mm sources are arguably resolvable. This is consistent with predicted resolution of 2.1\,mm when PSF is not corrected for. However, artifacts occur near the center. As discussed above, we attribute them as aliasing errors due to undersampling of this region.
By using 0.3\,mm pixels, resolution of the S image is arguably improved but the aliasing artifacts near the center also become stronger and additional ring artifacts appear.
It is worth noting that that these S images were obtained by using only 4 subsets in OSEM.
When more subsets were used, artifacts are even more conspicuous and dominate the resulting images.

The M0 images also have deteriorated image quality near the center and contain streak artifacts in sectors 1-3. We again postulate that they are due to undersampling.
If we disregard such deterioration and artifacts, the 1.8\,mm, 2.1\,mm, and 2.4\,mm sources are substantially better visualized than in the S images.
With 500 OSEM iterations, the 1.5\,mm sources are arguably resolvable if they are not masked by artifacts.
Again, this is consistent with the view that PSF correction improves image resolution up to the degree allowed by the sampling condition.

Figure~\ref{fig:noise-free-center-M} shows the resulting noise-free images when modulators were used.
The M0.5 images are similar to the M0 images.
The M1 image at 500 iterations can clearly resolve 1.5\,mm and larger sources, showing no apparent resolution degradation near the center.
While not as strong as in the S, M0 and M0.5 images, some streak artifacts can still be observed, especially in sectors 1 and 2. These suggest that aliasing errors are mitigated but not eliminated.
For the M2 and M4 images at 500 OSEM iterations, aliasing errors are absent.
Subjectively, the M2 images have the best resolution and the 0.9\,mm sources are arguably distinguishable.

\begin{figure}[t]
    \centering
    \includegraphics[width=1.0\linewidth]{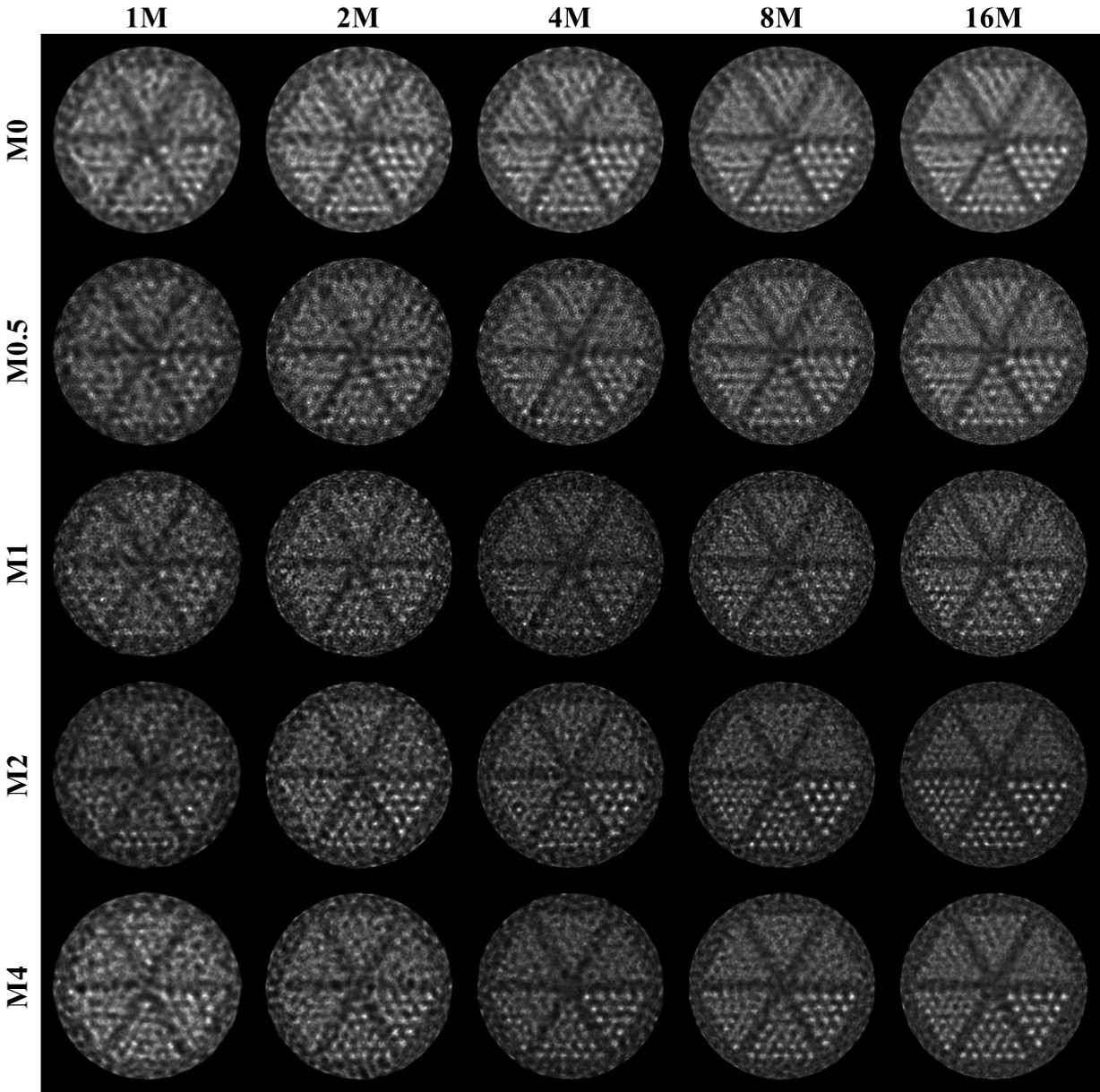}
    \caption{From left to right are images obtained from 1M-, 2M-, 4M-, 8M-, and 16M-event data by using 10, 20, 30, 40 and 50 iterations, respectively.}
    \label{fig:noisy}
\end{figure}

\begin{figure}[t]
    \centering
    \includegraphics[width=0.95\linewidth]{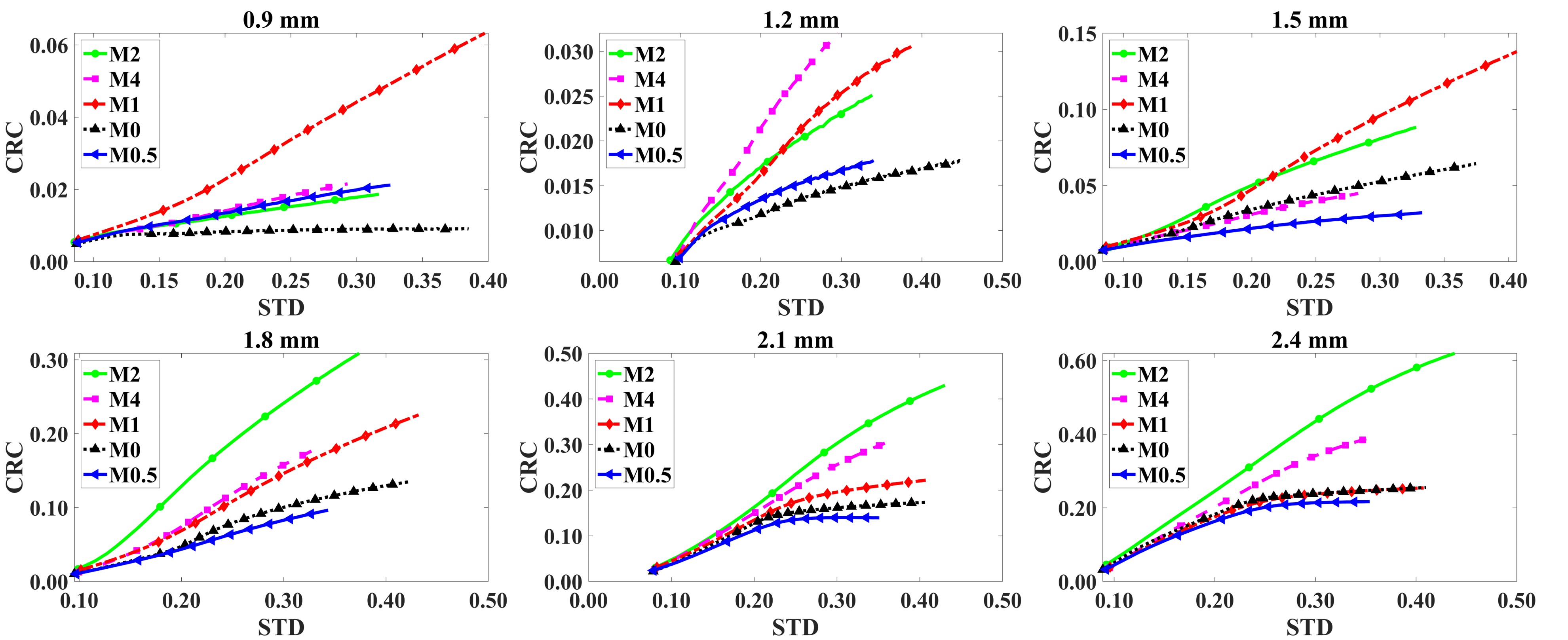}
    \caption{CRC-vs-STD tradeoff curves for various source sizes and modulators when increasing the OSEM iteration number from 1 to 150 for M0, and from 1 to 50 for others. Markers identify data points at every 5 iterations.}
    \label{fig:crc}
\end{figure}

Figure~\ref{fig:noisy} shows images obtained from analytically simulated noisy data containing 1 million (M) to 16M events, with the phantom placed at the center of the scanner.
As described above, images identified by the same event count are derived from data simulated for the same scan duration and the counts given are those of the M0 data.
Due to CDE reduction by the modulators, the actual counts for other M data are the counts shown multplied by $\beta(5\,\text{mm})=0.57$.
The number of iteration is subjectively chosen to yield a good balance between resolution and noise. Generally, and as expected, the image resolution and quality improve as the event count increases. Again, the M0, M0.5 and M1 images exhibit aliasing errors and the M2 images have the best resolution. Compared to M0, despite roughly 43\% reduction in CDE, M2 yields better contrast for sources larger than 1.8\,mm, even with 1M-event data.
With 16M-event data, the 1.2\,mm sources are arguably observable in M2.

\subsection{CRC-vs-STD trade-off}
Figure~\ref{fig:crc} shows the CRC-STD tradeoff curves obtained using the 8M-event data as the OSEM iteration number varies.
In this plot, at the same STD a curve having a larger/smaller CRC than the other is considered superior/inferior.
Compared to the M0 curve, the M0.5 curve is superior for 0.9\,mm and 1.2\,mm sources but is inferior for other larger sources;
the M1 curve is superior for 1.8\,mm and smaller sources and is comparable for 2.1\,mm and 2.4\,mm sources;
the M2 trace is superior for all source sizes;
and the M4 trace is superior for all source sizes but 1.5\,mm (becoming slightly inferior).
The M1 curve has the largest CRC for the 0.9\,mm sources.
However, as observed above, 1.5\,mm and smaller sources in the M0, M0.5, and M1 images (and some 1.8\,mm sources close to center) are affected by aliasing errors.
Therefore, this apparent superiority of M1 is likely to be an artifact.
Comparing M2 and M4, the latter is inferior to the former for 1.5\,mm and larger sources but is superior for smaller sources.
Overall, M2 consistently produces superior quantitative results for all source sizes.
Specifically, M2 can increase the largest CRC achieved by M0 by a factor of $\geq$     1.5 for 1.5\,mm and smaller sources, and by a factor of $\geq$2.3 for 1.8\,mm and larger sources, while maintaining the STD.

\subsection{Modulation depth}

 \begin{figure}
    \centering
    \includegraphics[width=0.9\linewidth]{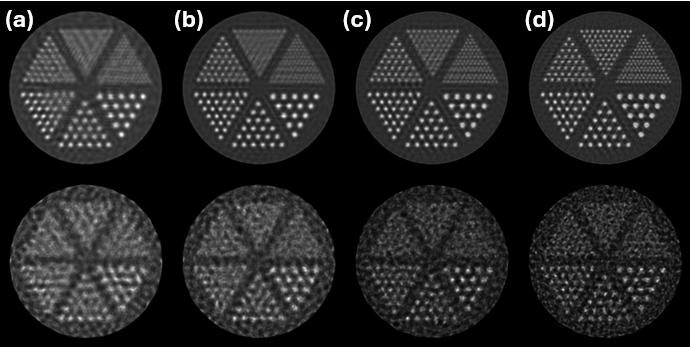}
    \caption{Images obtained for M2 employing 1\,mm (a), 2.5\,mm (b), 5\,mm (c) and 10\,mm (d) thickness tungsten, with the phantom 
    placed at the center of the scanner, by using 500 iterations of noise-free data (top) and 50 iterations of the 8M-event data (bottom).}
    \label{fig:value}
\end{figure}

\begin{figure}
    \centering
    \includegraphics[width=0.95\linewidth]{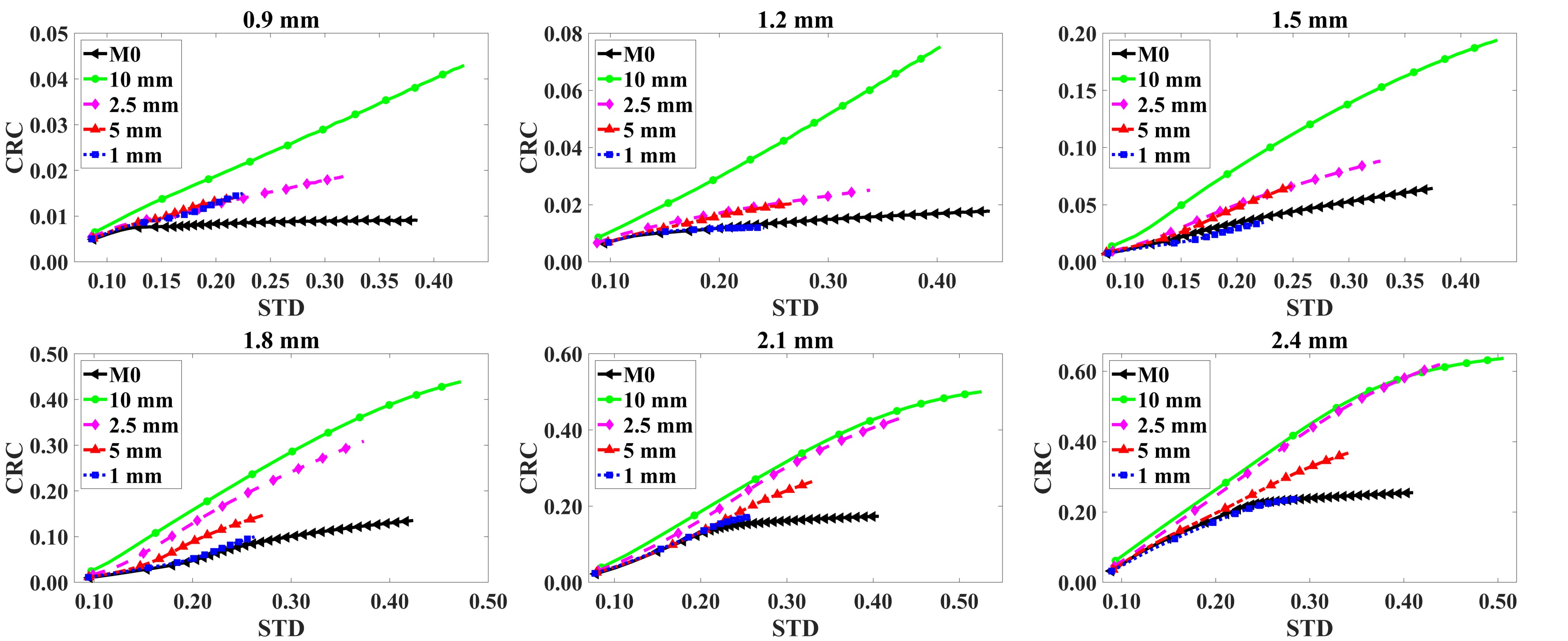}
    \caption{CRC-vs-STD tradeoff curves for 8M-event data when using the M2 modulator with 1\,mm, 2.5\,mm, 5\,mm and 10\,mm thickness of tungsten.}
    \label{fig:value-crc}
\end{figure}

Figure~\ref{fig:value} shows the images obtained when using the M2 modulator whose transmission profile corresponding to various tungsten thicknesses,
including 1\,mm, 2.5\,mm, 5\,mm, and 10\,mm.
The phantom was placed at the center of the scanner.
Images in the first row were obtained by 500 OSEM iterations of analytically simulated noise-free data, those in the second row by 50 OSEM iterations of analytically simulated noisy data for the same scan duration that yielded 8M events for M0 (hence, 0.85$\times$8M, 0.71$\times$8M, 0.57$\times$8M, and 0.48$\times$8M events for M2 with 1\,mm, 2.5\,mm, 5\,mm, and 10\,mm thicknesses). 
From the noise-free images, it can be observed that a larger modulation depth created by use of a thicker tungsten can lead to better resolution recovery.
For the noisy images, the 10\,mm-thickness image still exhibits subjectively the best resolution despite having the largest CDE reduction.
Compared to the 8M-event M0 image shown in figure~\ref{fig:noisy}, it also shows significantly better image resolution even though it has less than half the event count. 

Figure~\ref{fig:value-crc} shows the CRC vs. STD tradeoff curves obtained for the 8M-event data when using the M2 modulator with different tungsten thicknesses.
At this noise level, thicker tungsten yields superior CRC vs. STD tradeoff curves for all source sizes, despite larger CDE reduction.
The improvement in CRC is particularly evident for 1.5\,mm and smaller sources.

\subsection{Images of the resolution phantom-- GATE simulated data}
\label{subsection:GATE_resolution_phantom}
\begin{figure}[t]
    \centering
    \includegraphics[width=1.0\linewidth]{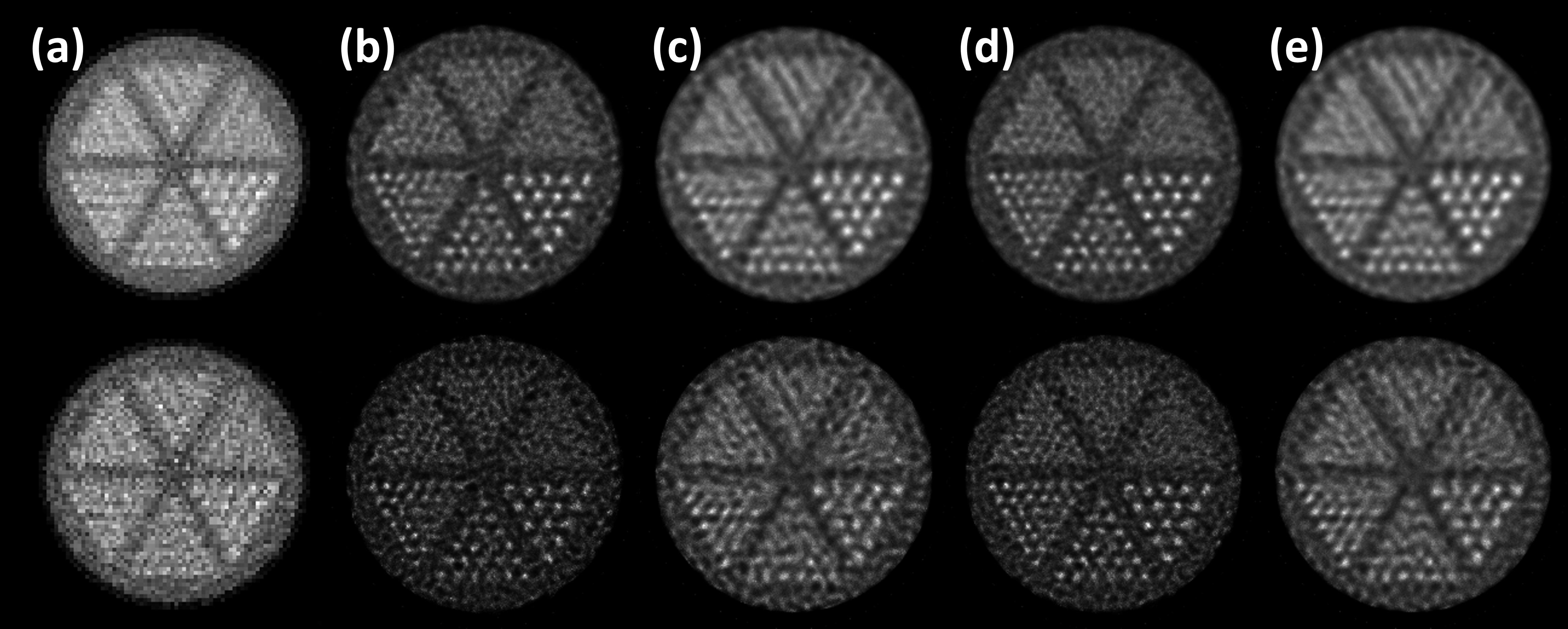}
    \caption{S (a), M2 (b,d), and M0 (c,e) images of the resolution phantom obtained from GATE simulation data.
    Two datasets corresponding to two different scan durations were produced for the situations when no modulator was used (S and M0) and when the M2 modulator was used.
    The M2 and M0 data in dataset 1 had 6.4M events (b) and 11.0M events (c), respectively.
    In dataset 2, they had 10.4M events (d) and 17.8M events (e).
    The S image used the M0 data in dataset 1.
    Images in the top (bottom) row were obtained using 5 (30) OSEM iterations for S and 30 (100) iterations for M0 and M2.}
    \label{fig:gate}
\end{figure}

Figure~\ref{fig:gate} shows the S, M0, and M2 images obtained from GATE data.
For S, 1.0\,mm pixels were used while 0.3\,mm pixels were used for M0 and M2.
The phantom was placed at the center of the scanner, and 5\,mm thickness tungsten was used for M2.
Two datasets corresponding to two different scan durations were produced.
The M2 and M0 data in dataset 1 had, respectively, 6.4M and 11.0M events.
In dataset 2, they had 10.4M and 17.8M events.
As with analytically simulated data, SR-PET was successfully achieved for GATE data.
Again, the S images have the worst resolution;
the S and M0 images exhibits streak artifacts;
and the M2 images have considerably better resolution than others.
For the two noise levels considered, the 1.8\,mm sources are resolvable in M2 but not in M0 when using 30 iterations.
Using 100 iterations for M2, resolution is further improved with the 10.4M-event data but noise is considerably amplified with the 6.4M-event data.
Compared to the images shown in figures~\ref{fig:noise-free-center-M0}-\ref{fig:noisy}, aliasing artifacts in S and M0 are not as conspicuous.
We postulate that this may be due to the wider PSFs in the case of GATE data.

\begin{figure}[t]
    \centering
    \includegraphics[width=0.95\linewidth]{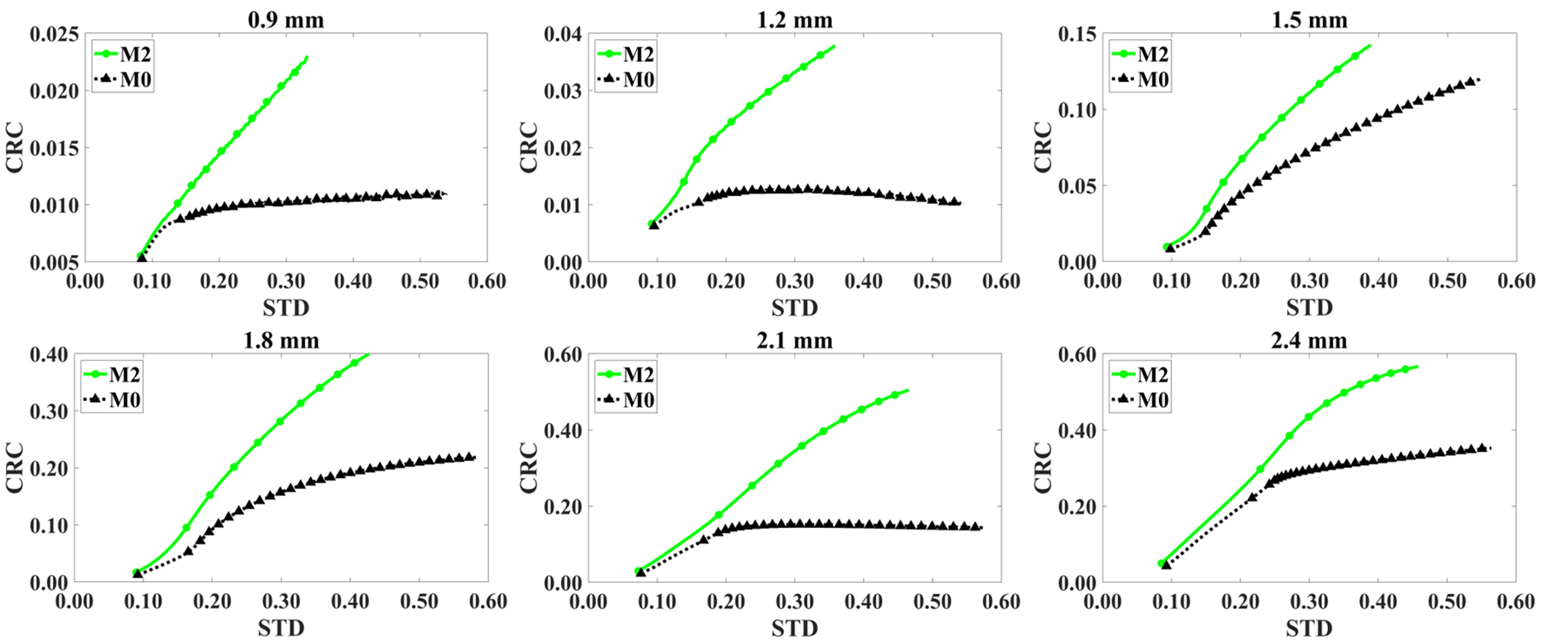}
    \caption{CRC-STD tradeoff curves obtained for the resolution phantom  from dataset 2 produced by GATE. Markers identify data points obtained at every 10 OSEM iterations.}
    \label{fig:CRC-gate}
\end{figure}

Figure~\ref{fig:CRC-gate} shows the CRC-STD tradeoff curves obtained using dataset 2.
At the same number of iterations, nearly every M2 data point lies above and to the right of the corresponding M0 point, indicating that M2 image has higher spatial resolution but also greater image noise compared to M0.
For all source diameters, the M2 curve is consistently above the M0, demonstrating that M2 image is quantitatively superior.

\subsection{Images of the brain phantom}
\label{subsect:Analytic-brain}
    \begin{figure}[t]
    \centering
    \includegraphics[width=0.67\linewidth]{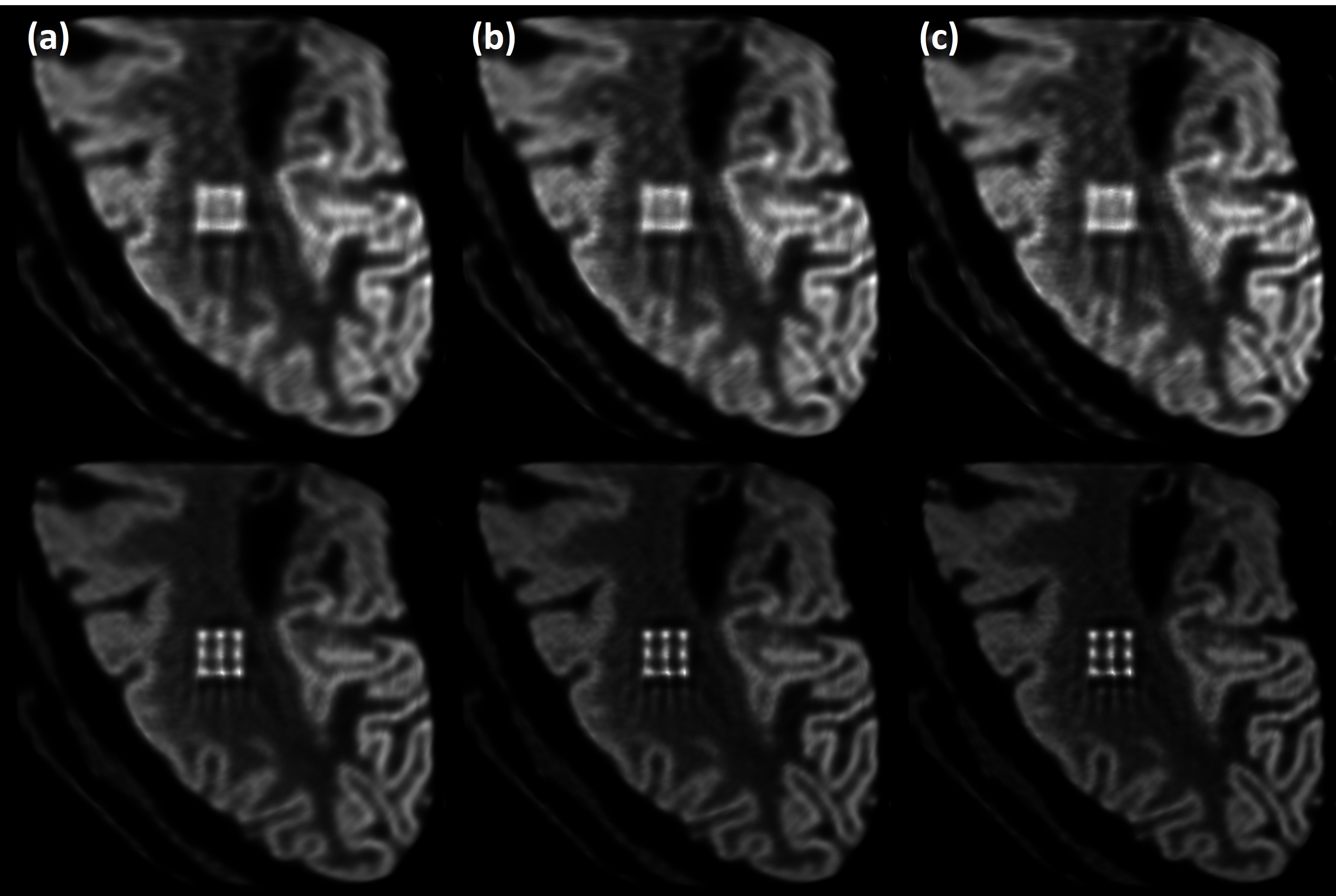}
    \caption{M0 (top) and M2 (bottom) images obtained from analytically simulated noise-free data for the brain phantom by using 50 (a), 100 (b), and 200 (c) OSEM iterations.}
    \label{fig:brain-1}
\end{figure}
Figure~\ref{fig:brain-1} shows M0 (top row) and M2 (bottom row) images obtained for the brain phantom by using various number of iterations when using analytically calculated noise-free data.
Again, the phantom was placed at the center of the scanner, and 5\,mm thickness tungsten was assumed for M2. 
At the same number of iterations, M2 has better resolution than M0.
The M2 image can clearly distinguish the positions and boundaries of the 9 lesions, while the M0 image fails to resolve them.
Additionally, the M2 image also shows better definition of the gray matter gyri.



    \begin{figure}
    \centering
    \includegraphics[width=1.0\linewidth]{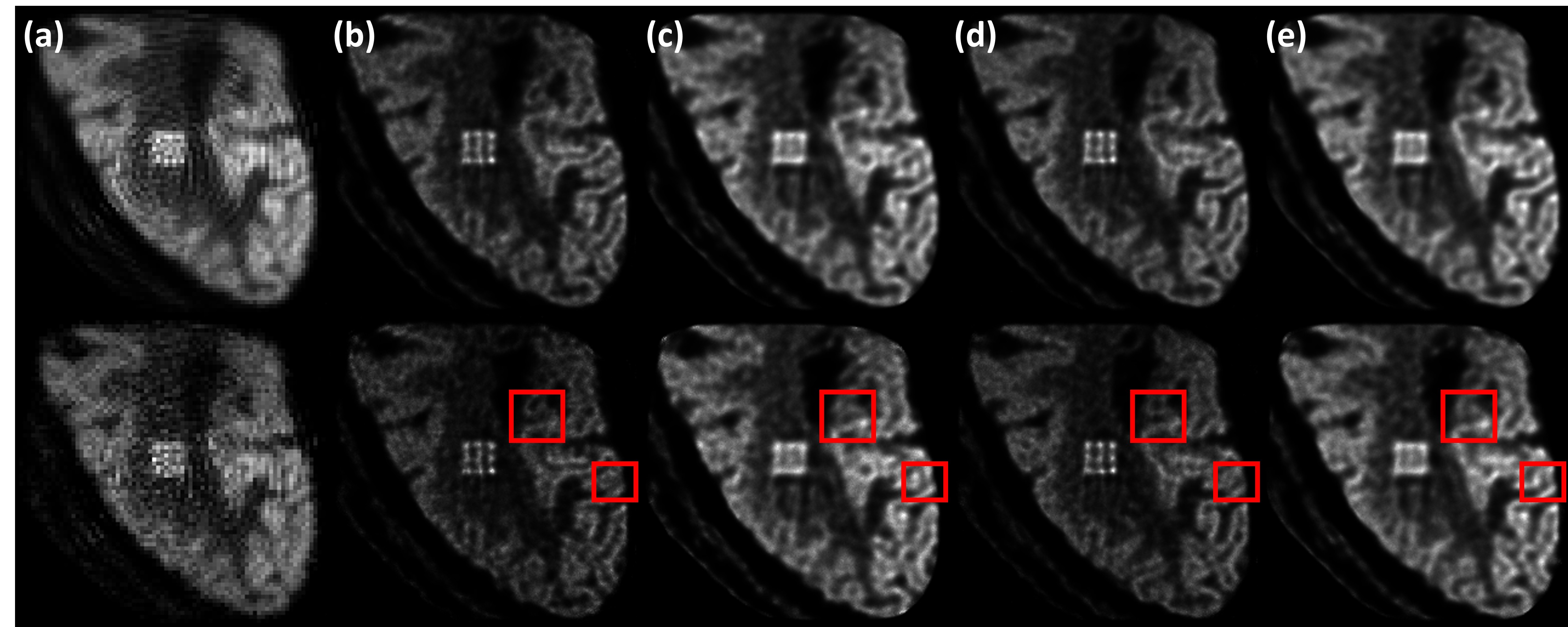}
    \caption{S (a), M2 (b,d), and M0 (c,e) images of the brain phantom obtained from GATE data.
    Two datasets corresponding to two different scan durations were produced for the situations when no modulator was used (S and M0) and when the M2 modulator was used.
    The M2 and M0 data in dataset 1 had 7.7M events (b) and 13.2M events (c), respectively.
    In dataset 2, they had 13.4M events (d) and 23.0M events (e).
    The S image used the M0 data in dataset 1.
    Images in the top (bottom) row were obtained using 25 (50) iterations.
    Compared to the phantom in figure~\ref{fig:phantom}(b), structural errors in the M0 images can be observed (the red rectangles show two examples).}
    \label{fig:brain-gate-2}
\end{figure}

Figure~\ref{fig:brain-gate-2} shows the S, M0, and M2 images obtained from GATE data for the brain phantom.
As in section~\ref{subsection:GATE_resolution_phantom}, 1.0\,mm pixels were used for S and 0.3\,mm  pixels were used for M0 and M2.
Two datasets simulating different scan durations were produced, yielding 7.7M and 13.2M events for M2 and M0 in dataset 1 and 13.4M and 23.0M events for M2 and M0 in dataset 2.
Again, it can be observed that the S image has the lowest spatial resolution and also suffers from ring-like artifacts.
Under the same scan duration and OSEM iteration number, even though it had only about 43\% fewer events,  M2 shows considerably better resolution than M0. 
For both datasets, the lesions are resolved in M2 but not in M0.
Visually, M2 can clearly resolve most gyri while the boundaries are less distinct in M0.
Some structural errors also appear in M0.


\section{Discussion and conclusion}
This paper has three main contributions.
The first is a theoretical exposition of the applicability of SR-SIM for increasing the resolution and eliminating aliasing errors from multiple undersampled bandlimited measurements.
The second is the idea of using a rotating modulator ring in front of a stationary PET detector ring to produce the needed signal modulation and the formulation of a mathematical model for extending the basic theory of SR-SIM to the resulting data for achieving SR-PET.
The third is numerical demonstration of the feasibility of this new SR-PET concept.
Using noise-free data for a 2-d PET system consisting of 4.2\,mm-width detectors produced by an analytic method, we show that the proposed SR-PET method can remove aliasing errors and resolve 0.9\,mm sources when using rotating bi-level ring modulators.
When Poisson noise is added, although the modulator decreases the CDE, improved visibility and superior CRC-vs-STD performance are still achieved.
Our results show that shorter-period modulators are not necessary for achieving higher resolution.
Among the four modulators tested, M2 gives the best resolution and quantitative performance, superior to M0.5 and M1.0.
This result is favorable because in practice longer-period modulators are easier to make.
We also examine using various tungsten thickness for M2.
Despite CDE reduction, for the noise levels examined small sources are better resolved when thicker tungsten is used to create larger signal modulation.
We also show successful SR-PET when applied to a brain-like activity distribution.
Results obtained using more realistic GATE simulation data are consistent with the above observations.

To the best of our knowledge, the idea of employing a rotating ring modulator for improving PET resolution has never not been proposed.
The results reported in this paper based on simulation data produced by analytic and MC methods constitute a strong proof-of-concept for this new approach.
However, the effects of many physical factors, including subject attenuation, scatter, random, and DOI blurring, need to be further investigated.
SR-SIM can only surpass resolution limit due to factors that are present after modulation is applied.
Consequently, in theory DOI blurring can be removed but the effects of positron range or photon acolinearity cannot. As mentioned in section~\ref{sec:introduction}, positron range is negligible with F-18 based tracers.
In current clinical systems, photon acolinearity leads to a resolution of about 1.8\,mm but it can be reduced by decreasing the diameter of the PET scanner.
In addition to CDE reduction, the modulator also can create scatter.
It is noted that this effect was included in GATE simulation to some extent.
In this paper, we consider bi-level modulators because they can be easily made. However, other designs may lead to a better balance between resolution enhancement and CDE reduction. Finally, while convenient the OSEM algorithm may not be optimal for solving equation~\eqref{eq:PET-SIM-patternl-1}. In particular, better noise regularization strategies are desired.

\section*{Appendix: Generalized Sampling Theorem}
Papoulis shows in \parencite{Papoulis1977GeneralizedSampling} that a $\sigma$-bandlimited signal $f(x)$ can be determined from samples of the responses of a number of LSI systems with input $f(x)$ obtained at a sub-Nyquist rate. Restating it in a form suitable for this paper, let $q_k(x)=r_k(x)\star f(x)$, or $Q_k(\nu)=R_k(\nu)F(\nu)$, $k=-K,\cdots,K$, be the outputs of $(2K+1)$ linear systems with input $f(x)$ where $r_k(x)$'s are the response functions. Let $W_k(\nu,x)$ satisfy
\begin{equation}
    \sum_{k=-K}^K R_k(\nu+\ell\nu_s)W_k(\nu,x) = e^{j2\pi \ell\nu_s x}
    \label{eq:Generalized-sampling-1}
\end{equation}
for $k,\ell=-K,\cdots,K$, where $\nu_s=2\sigma/(2K+1)$. Then,
\begin{equation}
    f(x) = \sum_{n=-\infty}^\infty \sum_{k=-k}^k q_k(n \delta x)w_k(x-n\delta x),
    \label{eq:Generalized-sampling-2}
\end{equation}
where $\delta x = 1/\nu_s$ and
\begin{equation}
    w_k(x) = \frac{1}{\nu_s} \int_{-\nu_s/2}^{+\nu_s/2} W_k(\nu,x)e^{j2\pi\nu x} d\nu.
    \label{eq:Generalized-sampling-3}
\end{equation}
The two necessary conditions are that equation~\eqref{eq:Generalized-sampling-1} is invertible and that the resulting solution allows equation~\eqref{eq:Generalized-sampling-3} to be evaluated.


\funding{This work was supported in part by the National Research and Development Program for Major Research Instruments of the Natural Science Foundation of China under Grant 62027808 and Grant 61927801; in part by the Sino-German Mobility Programme under Grant M-0387; and in part by NIH Grant R01 EB029948.}




\printbibliography

\end{document}